\documentstyle[aps,multicol,prl,epsfig]{revtex}
\begin{document}
\draft

% rp01.tex  CSOH       8/1/01
% rp02.tex  CSOH       9/4/01
% rp03.tex  AJL,SRN   10/22/01
% rp04.tex  AJL       10/23/01
% rp05.tex  CSOH      10/24/01
% rp06.tex  SRN       10/28/01
% rp07.tex  CSOH      10/29/01
% rp08.tex  everybody  10/29/01

\title{Random Packings of Frictionless Particles}
\author{Corey S. O'Hern$^{1,3}$, Stephen A. Langer$^2$, Andrea J. Liu$^1$,
and Sidney. R. Nagel$^3$}
\address{$^1$~Department of Chemistry and Biochemistry, UCLA, Los 
Angeles, CA  90095-1569}
\address{$^2$~Information Technology Laboratory, NIST,
Gaithersburg, MD 20899-8910}
\address{$^3$~James Franck Institute, The University of Chicago,
Chicago, IL 60637}
\date{\today}
\maketitle

\begin{abstract}
      We study random packings of frictionless particles at $T=0$.
      The packing fraction where the pressure becomes nonzero is the
      same as the jamming threshold, where the static shear modulus
      becomes nonzero.  The distribution of threshold packing
      fractions narrows and its peak approaches random close-packing
      as the system size increases.  For packing fractions within the
      peak, there is no self-averaging, leading to exponential decay
      of the interparticle force distribution.
\end{abstract}

\begin{multicols}{2}
\narrowtext

\begin{figure}
\epsfxsize=3.0in \epsfbox{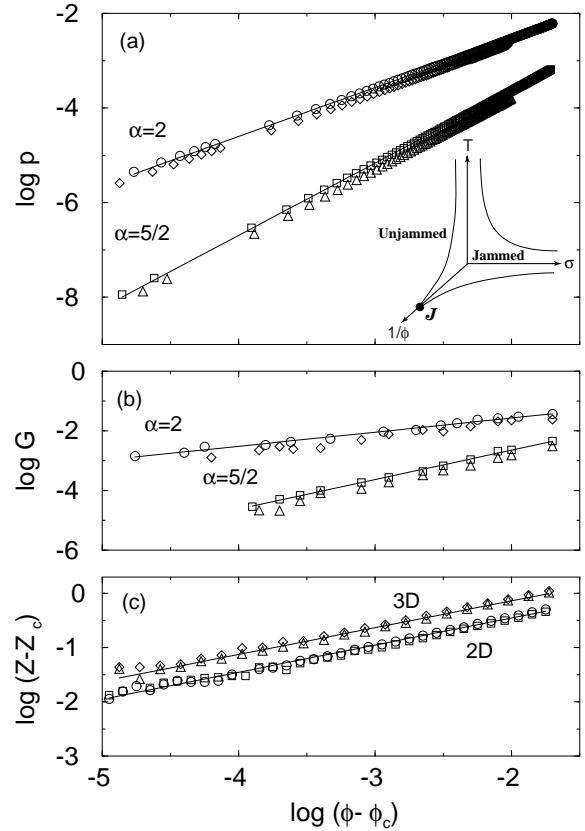}
\caption{(a) Pressure $p$ vs.
$\phi-\phi_c$, where $\phi_{c}$ is the threshold packing fraction for 
a given state.  Circles (squares)
correspond to $\alpha=2 (5/2)$ in $2D$ and diamonds (triangles)
correspond to $\alpha=2 (5/2)$ in $3D$ and $N=1024 (512)$ in
$2D (3D)$.  Inset: Jamming phase diagram showing onset of
jamming at point ${\rm J}$. (b) Shear stress $G$ vs. $\phi-\phi_c$.
The shear strain is applied in the $x$-direction,
the strain gradient is in the $y$-direction, and the $xy$ component of
the stress tensor is measured. (c) Number of overlaps per particle
$Z-Z_{c}$ vs. $\phi-\phi_c$.}
\label{pgz}
\end{figure}

A system jams when it develops a yield stress or extremely long stress
relaxation time in a disordered state\cite{book}.  Different control
parameters can be varied to induce jamming, such as the temperature
$T$, the applied shear stress $\sigma$, or the packing fraction $\phi$,
as shown in the phase diagram inset to Fig.~\ref{pgz}(a) \cite{lucid}.
Such a phase diagram might apply {\it e.g.\/} to supercooled liquids, granular
materials, foams and suspensions.  For the diagram to be
useful, there should be a common physical origin for jamming
independent of the control parameter varied.  Previously, it was
shown \cite{ohern} that a peak in the distribution of interparticle
normal forces, $P(F)$, signifies the development of a yield stress in
a variety of systems\cite{remarkxtal}, implying that the jamming phase
diagram is a useful concept.

There is a special point on the jamming phase diagram, marked ${\rm
J}$ in Fig.~\ref{pgz}(a), for repulsive, finite-range potentials.
This point, at zero temperature and zero shear stress, represents the
onset of jamming with increasing packing fraction.  Static granular
packings must necessarily lie near this point because they are
effectively at $T=0$ (the thermal energy is much smaller than the
energy needed to lift a grain by its own diameter) and the particles
are hard, so it is difficult to compress packings further into the
jammed region.  Experimentally, $P(F)$ for granular packings has a
remarkably robust form\cite{nagel}; not only does it have a peak, but
it also has an exponentially decreasing tail at large $F$.  Numerous
simulations find the same form for $P(F)$\cite{makse,thornton}.  The
persistence of the exponential tail, independent of the potential
used, is surprising.

In this letter, we examine configurations created by quenching systems
from high temperature to $T=0$ (and also $\sigma=0$) near the onset of
jamming (point ${\rm J}$ in the jamming diagram).  We find that
different configurations have different properties, even for
arbitrarily large system sizes, so that self-averaging is not
observed.  However, the range of packing fractions over which
self-averaging is not observed narrows with increasing system size. 
We will show that
one consequence of non-self-averaging is an exponential decay of
$P(F)$ at large $F$.

%%%%%%%%%%%%%%%%%
%%%%%%%%%%%%%%%%%

To create configurations near point ${\rm J}$ we start with random
configurations (i.e. $T=\infty$) and, as with inherent
structures\cite{stillinger}, quench infinitely rapidly to $T=0$ at
fixed $\phi$ using conjugate gradient (CG) energy
minimization\cite{numericalrecipes}.  We study $50:50$ binary mixtures
of $N$ frictionless particles with a diameter ratio $1.4$
\cite{speedy,perera} that interact pairwise via repulsive potentials:
$v_{ab}(r) = (\epsilon/\alpha)~(1 - r/\sigma_{ab})^\alpha$ for
$r<\sigma_{ab}$ and $v_{ab}(r) = 0$ for $r>\sigma_{ab}$, where $a,b$
label particles and $\sigma_{ab} = (\sigma_a+\sigma_b)/2$.  We study
harmonic ($\alpha=2$) and Hertzian ($\alpha=5/2$) potentials in $2D$
and $3D$. The total potential energy is $V=0$ if no particles overlap.
Energy is measured in units of $\epsilon$ and length in units of the
small particle diameter $\sigma_1$.

We classify each final configuration as either
overlapped ($V \ne 0$) or non-overlapped ($V=0$).  Overlapping
configurations must have a nonzero pressure, $p$,
while non-overlapping ones have $p=0$.  Is a configuration that has
$p>0$ necessarily jammed with a zero-frequency shear modulus $G>0$?
To answer this question, we study states close to overlap threshold as
a function of packing fraction.  We find the overlap threshold,
$\phi_{c}$, for each configuration by compressing a non-overlapped
state while measuring $p$.  Throughout the compression,
after each small increment in $\phi$ ($\delta \phi \le 10^{-4}$), we
use CG to
bring the state to the lowest energy attainable without crossing any
barriers.  (To check that no barriers were crossed, we reproduced our
results using ten times smaller increments in $\phi$).  This 
procedure allows us to measure
zero-frequency properties of the system.

Different states have different values of the overlap
threshold, $\phi_{c}$.  Nevertheless, when we plot pressure $p$ versus
$\phi-\phi_{c}$ (Fig.~\ref{pgz}(a)), the results for
different configurations collapse on a single
curve.  This holds for both harmonic
($\alpha=2$) and Hertzian ($\alpha=5/2$) potentials in $2D$ and $3D$. We
find $p = p_{0} (\phi-\phi_{c})^{\beta}$, where $p_{0}$ is only weakly
dependent on dimension and $\beta=1.0$ for harmonic and $\beta=1.5$
for Hertzian potentials, independent of dimension.  This is consistent
with previous results at larger $\phi-\phi_{c}$
\cite{makse,doug,lacasse}.  To see if states with $p>0$
are jammed, we calculate the shear modulus,
$G$, by applying a small step strain and measuring the infinite-time
response by minimizing the energy using CG. The response is
linear for sufficiently small strains.  Fig.~\ref{pgz}(b) shows that $G = G_{0}
(\phi-\phi_{c})^{\gamma}$, where $\gamma=0.5$ for $\alpha=2$\cite{doug},
and $\gamma=1.0$ for $\alpha=5/2$, in $2D$ and $3D$. To
our resolution, which is
better than $10^{-4}$, we find that $G$ and $p$ vanish at the
same packing fraction $\phi_{c}$.  This implies that the onset
of jamming, as defined in the first paragraph, coincides with the
onset of overlap\cite{cates}.

When a configuration jams at $\phi_{c}$, the number of overlaps per
particle, $Z$, jumps from zero to a threshold value $Z_{c}$.  Above
$\phi_{c}$, $Z$ increases as $Z-Z_{c}=Z_{0} (\phi-\phi_{c})^{\zeta}$,
as shown in Fig.~\ref{pgz}(c).  For both harmonic and Hertzian
potentials, we find $\zeta=0.5$ in both $2D$ and $3D$\cite{doug}.  (If
there are no zero-energy modes, $Z_{c}=2 d$ for frictionless spheres
in $d$ dimensions\cite{book}.  Usually $\approx 5\%$ of particles are
``rattlers'' that do not overlap with any neighbors.  If we remove
these, we find $Z_{c}=2 d$.)

By studying the onset of jamming in repulsive systems at $T=0$, we have
a criterion for whether a state is jammed or not ({\it i.e.} $V>0$ or
$V=0$).  If there is a packing with $V=0$, then it is an
equilibrium configuration.  The state that maintains $V=0$ up to the
highest $\phi$ is therefore the one that, when compressed
infinitesimally above this point, becomes the zero-temperature
equivalent to the ideal glass.  We have found that the properties
shown in Fig.~\ref{pgz} depend only on $\phi-\phi_{c}$.  This suggests
that this behavior is the same as for the ideal glass.

%%%%%%%%%%%%%%%%%
\begin{figure}
\epsfxsize=3.2in \epsfbox{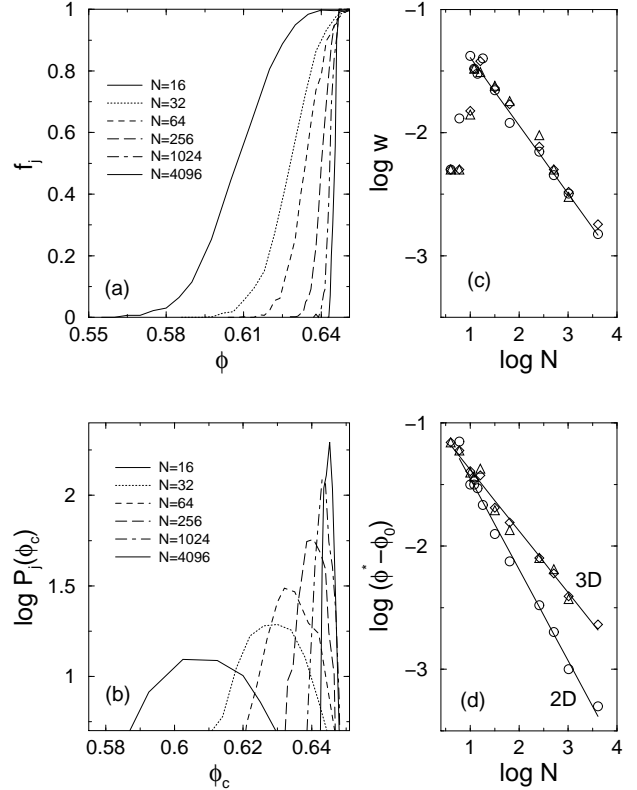}
\caption{(a) The probability $f_j$ of finding a jammed state vs. $\phi$
for $\alpha=2$ in $3D$ and several system sizes, $N$.  (b)
The probability distribution $P_j(\phi_c)$ of finding a jamming threshold
$\phi_c$ for systems considered in (a).  (c) Full-width at half-maximum
$w$ of $P_j(\phi_c)$ vs. $N$. (d) The
deviation in peak position $\phi_0$ of $P_j(\phi_c)$ from
its asymptotic value $\phi^{*}$ vs. $N$.  Data for (c) and (d) are for
systems considered in Fig.~\ref{pgz} except $\alpha=5/2$ in $2D$.}
\label{mrj}
\end{figure}
%%%%%%%%%%%%%%%%%

Although Fig.~\ref{pgz} shows scaling behavior above
$\phi_{c}$, the value of $\phi_{c}$ varies from state to state.  How
much can $\phi_{c}$ vary?  Our protocol, of starting with random
configurations and quenching infinitely rapidly to $T=0$
should, in principle, allow us to sample all of
phase space.  For each $\phi$, we measure the
fraction of initial states that lead to jammed states at $T=0$.
Fig.~\ref{mrj}(a) shows
the probability, $f_{\rm j}(\phi)$, of finding a
jammed state for
different system sizes in $3D$ for the harmonic potential.  For
each $N$, we have differentiated $f_{\rm j}$ with respect to
$\phi$ to obtain the probability distribution $P_{\rm j}(\phi_{c})$ of
finding a jamming threshold of $\phi_{c}$ (see Fig.~\ref{mrj}(b)).
System size has a large effect.  For each $N$, we
characterize $P_{\rm j}(\phi_{c})$ by
its full width at half maximum, $w$, and its peak position,
$\phi_{0}$.  Fig.~\ref{mrj}(c) shows that for $N>10$, $w \propto
N^{-\omega}$, where $\omega \approx 0.55$.  Similarly,
Fig.~\ref{mrj}(d) shows that $\phi_{0}$ approaches its asymptotic
value $\phi^{*}$ as $\phi^{*}-\phi_{0} \propto N^{-\theta}$, where
$\theta \approx 0.7$ for $2D$ and $\theta \approx 0.5$ for $3D$.  We find 
$\phi^{*}=0.842$ in $2D$ and
$\phi^{*}=0.648$ in $3D$, close to well-known values of random
close-packing\cite{rcp}.

For finite $N$, the peak $\phi_{0}$ of the distribution $P_{\rm
j}(\phi_{c})$ corresponds to the largest number of initial states that
lead to final jammed configurations (not the largest number of
distinct final jammed states).  As such, $\phi_{0}$ is a measure of
the largest fraction of phase space that leads to the onset of jamming
for a given $N$.  $\phi^{*}$ thus represents where the jamming
threshold is maximally random in the $N\rightarrow \infty$ limit.  We
find the same limiting value of $\phi^{*}$ for Hertzian and harmonic
potentials, suggesting that $\phi^{*}$ is not sensitive to the
potential.  We can approach jammed hard-sphere packings by noting that
states up to the jamming threshold are accessible to hard spheres.  If
we repeat the measurement of $\phi^{*}$ for potentials with
progressively harder repulsions, we can approach (but not attain) the
hard-sphere limit.  This suggests a way to measure the maximally
random jammed packing fraction for hard spheres\cite{torquato}.

It is well known that spherical granular materials can exist over a
15\% spread of packing fractions ranging from 0.55 to 0.64
\cite{bernal,onoda}.  The width of our distribution $P_{j}(\phi_{c})$
for isotropic packings of frictionless soft spheres is a maximum near
$N=10$ and becomes arbitrarily small in the $N\rightarrow \infty$
limit.  Only for $N \approx 10$ can we find states that are
jammed at packing fractions as low as 0.55.  Thus, the experimental
difference between loose- and close-packing cannot be explained
simply by considering allowed configurations of frictionless
spheres; the value of loose-packing is not a purely geometrical
quantity.

We note that there are other protocols for finding the jamming
threshold at $T=0$.  We have also generated configurations by cooling
slowly from equilibrium thermal states to $T=0$.  In that case, we
find that $P_{j}(\phi_{c})$ is shifted to higher values of $\phi_{c}$,
with values of $\phi_{0}$ that are less than 1\% higher than for the
first protocol, but it is difficult to determine whether the
difference persists when $N\rightarrow \infty$.

We have shown that as $N \rightarrow \infty$, the range of packing
fractions over which systems can jam becomes arbitrarily narrow.
Therefore, one might suppose that the fact that different
configurations jam at different packing fractions might become
irrelevant in the large $N$ limit.  This is not the case because
within the range over which both jammed and unjammed configurations
exist, there is no self-averaging.  For example, if a configuration is
unjammed, with $V =0$, every subset of that configuration has $V=0$,
as well.  Less obvious is that for jammed configurations,
all subsets of more than a few particles will likely also contain overlaps.
This is found numerically, and stems from the fact that on average,
each jammed particle must have at least $2d$ overlapping contacts, each of
which also has $2d$ contacts.  This constraint makes it unlikely that a
jammed system can exist with pockets containing more than a few rattlers.

%%%%%%%%%%%%%%%%%
\begin{figure}
\epsfxsize=3.2in \epsfbox{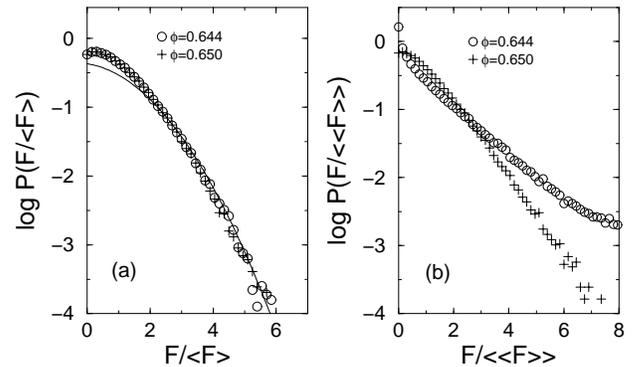}
\caption{(a) Force distribution $P(F/\langle F\rangle)$ obtained by
scaling $F$ by the average force $\langle F\rangle$ of each
configuration using the harmonic potential in $3D$ with $N=1024$.
The solid line is a Gaussian fit to the
high-force tail.  (b) Same as (a) except $F$ is scaled by $\langle
\langle F\rangle \rangle$ averaged over all configurations.}
\label{pf}
\end{figure}
%%%%%%%%%%%%%%%%%

The distribution of interparticle normal forces, $P(F)$, illustrates
the absence of self-averaging.  Simulations of static
packings\cite{makse,thornton} have shown that at large $F$, $P(F)$
falls off exponentially even for harmonic and Hertzian potentials.  An
argument based on equilibrium liquids\cite{ohern}, however, suggests
that the distribution should fall off differently for different
potentials: the large force tail depends on the pair-distribution
function at small $r$, which in equilibrium varies approximately as
$\exp(-V(r)/kT)$.  By this reasoning, one would predict a Gaussian
tail for $P(F)$ for a harmonic potential.  Indeed, we do find a
Gaussian tail except at $T=0$ near $\phi_{0}$ (the peak of
$P_{j}(\phi_{c})$), where both jammed and unjammed configurations can
exist.  In this region, we obtain a different result for $P(F)$,
depending on whether we scale $F$ by its average value before or after
taking the configurational average.  If, for each configuration, the
forces $F$ are scaled by the average value {\it for that
configuration}, $\langle F\rangle$, and the resulting distribution
$P(F/\langle F\rangle)$ is then averaged over all configurations, the
result is as shown in Fig.~\ref{pf}(a).  On the other hand, if the
forces are scaled by the average over {\it all} configurations,
$\langle\langle F\rangle\rangle$, the resulting $P(F/\langle\langle F
\rangle \rangle)$ is as shown in Fig.~\ref{pf}(b).  We see that at
$\phi=0.644$, near the peak of $P_{j}(\phi_{c})$ for $N=1024$, the
high force tail falls off much more slowly in this latter case than in
the former, where each configuration is averaged separately.  As the
packing fraction is increased above $\phi_c$, there is less difference
between the curves $P(F/\langle\langle F\rangle\rangle)$ and
$P(F/\langle F\rangle)$.  This trend can be explained since near
$\phi_c$, $\langle F\rangle$ varies dramatically from configuration to
configuration.  The relative size of fluctuations in $\langle
F\rangle$ decreases with increasing $\phi$, and the exponential decay
of $P(F)$ thus disappears.  As the size of the system increases, the
width of the region in $\phi$ over which the high-force tail is
exponential will decrease.  However, one can always tune $\phi$ close
enough to $\phi_{0}$ to observe the exponential tail.

The shape of the tail of $P(F/\langle\langle F\rangle\rangle)$ can be
computed analytically, given a few simple assumptions.  We have found
that the average force in a configuration, $\langle F \rangle$, is
proportional to the pressure $p$ at $T=0$ near the onset of jamming.
So from Fig.~\ref{pgz}(a), we know that $\langle F\rangle = F_0 (\phi
- \phi_c)$ for the harmonic potential.  We assume that the jamming
threshold, $\phi_{c}$, is distributed as a Gaussian centered at
$\phi_0$ with width $w$ as found in Fig.~\ref{mrj}(b).  Finally, we
assume (as shown in Fig.~\ref{pf}(a)) that the tail of $P(F/\langle F
\rangle)$ for individual configurations is given by the equilibrium
argument, which implies a Gaussian tail centered at $F/\langle F
\rangle=0$ with width $\sigma_{F}$.  (Here, $F_0$, $\sigma_F$,
$\phi_0$, and $w$ are parameters that can be obtained from the
simulation data.)  In the large $F$ limit, we find
\begin{eqnarray}
P({F}) & \propto & \int_{0}^{\phi} d\phi_{c} {1 \over \langle
F \rangle} e^{-F^{2}/(2 \langle F \rangle^{2} \sigma_{F}^{2})}
e^{- (\phi_{c}-\phi_{0})^{2}/(2 w^{2})} \\
& \approx & {\exp(-F/\langle\langle
F \rangle\rangle) \over \sqrt{F/\langle\langle
F \rangle\rangle}}.
\label{asymp}
\end{eqnarray}

We have also studied Hertzian potentials and find results similar to
those for the harmonic potential shown in Fig.~\ref{pf} and
Eq.~\ref{asymp}.  In experimental granular systems, where the
interparticle potential is expected to be Hertzian at contact, $P(F)$
has an exponential tail even for a single configuration\cite{nagel}.
We speculate that this is due to friction in the laboratory system,
which allows heterogeneities from region to region within a single
sample.

We have shown that for a finite-size system the jamming
phase diagram looks somewhat different from the one sketched in the
inset to Fig.~\ref{pgz}(a).  Instead of a well-defined point ${\rm
J}$, we find that there is a region of $\phi$, centered around
$\phi_{0}$ with width $w$, in which both jammed and unjammed states
can exist.  As the size of the system increases, this region shrinks
to the point ${\rm J}$.  Effects not included in our simulations, such
as the presence of friction, non-spherically symmetric potentials, or
anisotropic packing (such as sequential packing under gravity) may
prevent this region from disappearing.

In some ways, point ${\rm J}$ in the phase diagram resembles a
critical point: there is power-law scaling (Fig.~\ref{pgz}),
$P(F/\langle\langle F\rangle\rangle)$ has a robust exponential tail
independent of potential, and configurations are not self-averaging.
In the context of foam, it has also been speculated that point ${\rm
J}$ corresponds to rigidity percolation\cite{rigid}.  However, near
${\rm J}$ the behavior differs from ordinary critical behavior, where
configurations are not self-averaging once the correlation length
exceeds the system size.  There are no fluctuations near ${\rm J}$;
that is, an unjammed (jammed) configuration will be unjammed (jammed)
everywhere.  This, as well as the fact that no bonds exist at packing
fractions below ${\rm J}$, makes this transition also different from
rigidity percolation.  Moreover, even though we find power-laws near
${\rm J}$, there is a jump from $Z=0$ to $Z=Z_{c}$ and the exponents
depend on the potential but not on dimension.  Thus, point ${\rm J}$
has rather special properties.

We thank J.-P. Bouchaud, P. Chaikin, D. Durian, G. Grest, D. Levine,
J. Socolar, and T.  Witten for helpful discussions.  We acknowledge
support from NSF-DMR-0089081 (CSO,SRN) and DMR-0087349 (CSO,AJL) and
computing resources from the High Performance Computing Research
Facility at Argonne National Lab.

\end{multicols}

\begin{references}
\bibitem{remarkxtal} The peak in $P(F)$ is
more generally indicative of a yield stress, and appears also in crystals
and $1D$ chains.
\bibitem{cates} This result suggests that states at $\phi_{c}$ are not
``fragile'', as in M. E. Cates, J. P. Wittmer, J.-P. Bouchaud, and
P. Claudin, Phys. Rev. Lett. {\bf 81}, 1841 (1998).  It is possible
that states below $\phi_{c}$ can jam when a shear stress is applied.
Such states can support a pressure but are fragile because they will
yield when the shear stress is reversed.
\end{references}
\end{document}